\documentclass[vecphys]{svmult}

\usepackage{makeidx}         
\usepackage{graphicx}        
\usepackage{multicol}        
\usepackage[bottom]{footmisc}

\makeindex             

\def\frac#1#2{{\textstyle{{#1}\over {#2}}}}

\def\lsim{\mathrel{\rlap{\lower4pt\hbox{\hskip1pt$\sim$}}
    \raise1pt\hbox{$<$}}}
\def\gsim{\mathrel{\rlap{\lower4pt\hbox{\hskip1pt$\sim$}}
    \raise1pt\hbox{$>$}}}
\def\sqr#1#2{{\vcenter{\vbox{\hrule height.#2pt
         \hbox{\vrule width.#2pt height#1pt \kern#1pt
         \vrule width.#2pt}
         \hrule height.#2pt}}}}

\def\beq{\begin{equation}}
\def\eeq{\end{equation}}
\def\beqa{\begin{eqnarray}} 
\def\eeqa{\end{eqnarray}}

\def\laq{\raise 0.4 ex \hbox{$<$}\kern -0.8 em\lower 0.62 ex\hbox{$\sim$}}
\def\gaq{\raise 0.4 ex \hbox{$>$}\kern -0.7 em\lower 0.62 ex\hbox{$\sim$}}

\begin{document}

\title*{The adventures of Spacetime}
\author{Orfeu Bertolami}
\institute{Instituto Superior T\'ecnico, Departamento de F\'\i sica, \\
Av. Rovisco Pais 1, 1049-001, Lisboa, Portugal \\
\texttt{orfeu@cosmos.ist.utl.pt}}

\maketitle

\section{Introduction}
\label{sec:1}

Since the XIX century, it is known, through the work of Lobatchevski, 
Riemann and 
Gauss, that spaces do not need to have a vanishing curvature. 
This was for sure a revolution on its own, however, from the point of view of 
these mathematicians, the space of our day to day experience, 
the physical space, was still an essentially
{\it a priori} concept that preceded all experience and was independent of 
any physical phenomena. Actually, that was also the view of Newton and Kant with 
respect to time even though, for this two spacetime explorers, the world was Euclidean. 

As is well known, Leibniz held a very different opinion, 
as for him space and time 
were meaningless concepts if it were not for their relation with the material 
world. Starting with the concepts of space and time as 
quantities intrinsically related to matter, Hertz developed, between 1889 
and 1894, a new formulation of 
mechanics, which culminated in the posthumous publication in 1894 of the 
book {\it Die Prinzipien der Mechanik in neuem Zusammenhange dargestellt}. 
But of course, it was only through the General Theory of Relativity, in 1915, that it 
was understood that spacetime cannot be considered independently of matter at all. 
  
Stepping a bit back wards, it was through Special Relativity that it was understood that the independence of the 
laws of physics in inertial frames from the velocity of the frame of reference requires that space and time are treated 
on same foot. It was the mathematician Hermann Minkowski who, in 1908, realized that the unity of the laws of physics 
could be more elegantly described via the fusion of space and time into the concept of {\it spacetime}. 
Hence, the spacetime is 4-dimensional. For Minkowski 
however, the similarity between space and time was not complete, as he defined the time coordinate 
using the imaginary unity so to preserve the Euclidean signature of the spacetime metric. This description is not, as 
we know today, very satisfactory; one rather uses the Lorentzian signature for the spacetime metric.

Not much later, in 1909, the finish physicist Gunnar Nordstr\"om speculated that spacetime could very well have more 
than four dimensions. A concrete realization of this idea was put forward by Theodor Kaluza in 1919 and Oskar Klein in 
1925, who showed that an unified theory of gravity and electromagnetism could be achieved through a 5-dimensional version 
of General Relativity and the idea that the extra dimension was compact and very small, and could hence have passed undetected. 
This idea was very dear to Einstein, and this lead has been widely followed in further attempts to unify all known 
four interactions of nature. These developments, and most particularly General Relativity, 
represented a fundamental departure from the way XIX century 
mathematicians viewed space and also 
changed the attitude of physicists with respect to the physical world. Spacetime is not a passive 
setting for physics as it is the solution of the field equations for the gravitational field for a given matter distribution, 
and the former evolves along with spacetime. This methodology led physicists to describe nature along the lines 
of C\'ezanne's principle, that is {\it through the cylinder, the sphere, the cone ...}, i.e. through a geometrical or metrical 
description. Moreover, research in physics is now closely related with 
the ``spacetime adventures'' as, depending on the imposed conditions, spacetime can expand, shrink, 
be torn, originate ``baby'' spacetimes and so on. And it is 
through physics that spacetime acquires quite specific features. Let us 
introduce some examples.

The requirement of chiral fermions in four dimensions 
demand that, if there exist more than $4$ spacetime dimensions, then the total number of dimensions, $d$, must be even if all 
extra ones are compact \cite{Witten1}. To obtain a consistent effective 4-dimensional model arising from a $d$-dimensional 
Einstein-Yang-Mills theory, one should consider multidimensional universes of the form $M_d = {\bf R} \times G^{{\rm ext}}/H^{{\rm ext}} 
\times G^{{\rm int}}/H^{{\rm int}}$, where $G^{\rm ext(int)}$ and $H^{{\rm ext(int)}}$ are
respectively the isometry groups in 3($d$) dimensions. This technique, known as {\it coset space dimensional reduction} \cite{HM} 
(see Ref. \cite{KMRV} for an extensive discussion), is quite powerful and has been used in various branches 
of theoretical physics. 
In cosmology, when considering homogeneous and isotropic models (a 1-dimensional problem) it can be used, 
for instance, to obtain effective models 
arising from $4$-dimensional \cite{BMPV} and $d$-dimensional Einstein-Yang-Mills-Higgs theories 
\cite{BKM}. For the latter case, one considers 
for instance, $G^{\rm ext(int)} = SO(4)~(SO(d+1))$ and $H^{{\rm ext(int)}} = SO(3)~(SO(d))$ as
the homogeneity and isotropy isometry groups in $3$($d$) dimensions.

The demand that supersymmetry, a crucial property of the $10$-dimensional superstring theory, 
is preserved in $4$ dimensions requires that $6$ dimensions of the world are compact, have a complex structure,
no Ricci curvature and an $SO(3)$ holonomy group. That is, this compact space must correspond to a Calabi-Yau manifold 
\cite{CHSW}. Connecting all string theories through $S$ and $T$ dualities suggests the existence of an 
encompassing theory, M-theory, and that spacetime is $11$-dimensional \cite{Witten5}.
In another quantum approach to spacetime, loop quantum gravity, it is suggested that spacetime has, at 
its minutest scale, presumably the Planck scale, $L_P \simeq 10^{-35}$ m, 
a discrete structure \cite{ARSmolin}.

Actually, the discussion on the number of space ($n$) and time ($m$) dimensions is not a trivial one, as it is 
related with the predictive power of solutions of the 
partial differential equations (PDEs) that describe nature. 
Indeed, given the importance of second-order PDEs for physics, it is natural to draw some general conclusions about this type of 
PDEs \cite{Tegmark}. Consider a second-order PDE in ${\bf R}^d$, with $d = n+m$:

\beq
\left[\sum_{i=1}^{d} \sum_{j=1}^{d} A_{ij} {\partial \over \partial x_i}{\partial \over \partial x_j} +  
\sum_{i=1}^{d} v_{i} {\partial \over \partial x_i} + f\right]u = 0~,
\eeq
where the matrix $A_{ij}$, which can be taken without loss of generality to be symmetric, vector $v_i$ 
and the function $f$ are 
differential functions of $d$ coordinates. Depending on the signs of the eigenvalues of $A_{ij}$, 
the PDE is said to be:

\vspace{0.3cm}

\noindent
i) Elliptic in some region of ${\bf R}^d$, if all eigenvalues are negative or all positive.

\vspace{0.3cm}

\noindent
ii) Hyperbolic in some region of  ${\bf R}^d$, if one eigenvalue is negative and the remaining ones are positive (or vice-versa).

\vspace{0.3cm}

\noindent
iii) Ultrahyperbolic in some region of  ${\bf R}^d$, if at least two eigenvalues are negative and 
at least two are positive.

\vspace{0.3cm}

The crucial issue about PDEs is that only hyperbolic equations allow for a well-posed {\it boundary value problem}, that is, 
for an unique solution, and a  
well-posed {\it initial value problem}, that is, initial 
data that lead to future predictions on regions beyond the boundary data, excluding singular points. 
Elliptic PDEs, on the other hand, allow for a well-posed boundary value problem, but an ill-posed initial value problem, 
so that no predictions about the future on regions beyond the boundary data, that is beyond local observations, 
can be made.
Ultrahyperbolic PDEs, on their turn, have, for both space-like and time-like directions in a hypersurface, 
an ill-posed initial-value problem.

Hence, one sees that if 
$n=0$ for any $m$ or $m=0$ for any $n$, the resulting PDEs are elliptic and hence no predictions can be made. If, on the other hand, 
$m \ge 1$ and $n \ge 1$, the PDEs are ultrahyperbolic and hence lead to unpredictability.  

One can advance with other reasons for excluding certain combinations of $m$ and $n$. For instance, in a world where 
$n <3$, there is no gravitational force in General Relativity \cite{Deser}. Moreover, one should 
expect weird ``backward causality'' if $m > 1$. It has has been pointed long ago by Ehrenfest 
\cite{Ehrenfest}, that if $n > 3$, neither atoms nor planetary orbits can be stable. This feature is associated 
with the fact that solutions of the Poisson equation give rise to electrostatic and gravitational potentials for a point-like 
particle that 
are proportional to $r^{2-n}$ for $n>2$ and to forces that are proportional to $r^{1-n}$. For $n>3$ the two-body problem 
has no stable orbit solutions. The conclusion is that the 
choice $m=1$ and $n=3$ has quite desirable features and would be the selected 
one if one has reasons to think that the dimensionality of the world is chosen by selection arguments. 
We shall return to the issue of selection of ``worlds'' later on.

In what follows, we shall elaborate on how contemporary high-energy physics has changed our view of 
the spacetime structure; however, before that we 
shall make a detour and discuss some mathematical properties of spaces.

\section{Mathematical Spacetime}
\label{sec:2}

The historical development of the general theory of curved spaces, Riemannian geometry, has been guided and strongly influenced 
by the General Theory of Relativity. It is relevant to stress that, besides its mathematical interest, General Relativity 
has passed all experimental tests so far and is believed to be a sound description of the 
spacetime dynamics \cite{Will, Bertolami}. 

Let us present some of the basic features of mathematical space in view of their relevance to physics. 
A $d$-dimensional differentiable manifold $M$ endowed with a symmetric, non-degenerate second-rank tensor, the metric, $g$, 
is called a pseudo-Riemannian manifold, $(M, g)$. A pseudo-Riemannian manifold whose metric has signature $(+, ..., +)$ is 
said to be Riemannian. The metric of a pseudo-Riemannian manifold has a Lorentzian signature 
$(-, +, ..., +)$. A condition for a differentiable 
manifold to admit a Lorentzian signature is that it is noncompact or has a vanishing Euler characteristic.
A well known theorem due to the mathematician Tulio Levi-Civita, states that a pseudo-Riemannian manifold has a unique symmetric 
affine connection compatible with the metric, being hence equipped with geodesics.

Some spaces are of particular importance for physics, since they correspond to solutions of the 
Einstein equations with a cosmological 
constant\footnote{We use units where $c=\hbar = k=1$.}, $\Lambda$:

\beq
R_{\mu \nu} - {1 \over 2}  g_{\mu \nu} R  = 8 \pi G T_{\mu \nu} + \Lambda g_{\mu \nu}~,
\eeq
where $R_{\mu \nu}$ is the Ricci curvature of $M$, $R$ its trace, $G$ is Newton's constant 
and $T_{\mu \nu}$ is the energy-momentum tensor of matter in $(M, g)$. A minimal list includes:

\vspace{0.3cm}

\noindent
1) The {\it de Sitter} (dS) space\footnote{After the dutch astronomer Willem de Sitter, who in 1917 first described this space.} 
which corresponds to the $(d+1)$-dimensional hyperboloid 

\beq
-(x^0)^2 + (x^1)^2 + ... + (x^{d+1})^2 = r_0^2
\eeq
in a $(d+1)$-dimensional Minkowski space, which for the 
arbitrary constant $r_0$, satisfies the vacuum Einstein equations with a 
cosmological constant

\beq
\Lambda = {d(d-1) \over 2 r_0^2}~.
\eeq  

\vspace{0.3cm}

\noindent
2) The {\it anti-de Sitter} (AdS) space which corresponds to the universal cover of the $(d+1)$-dimensional hyperboloid, that is

\beq
(x^1)^2 + ...  + (x^d)^2 - (x^{d+1})^2 - (x^{d+2})^2 = - r_0^2
\eeq
in a $(d+2)$-dimensional space, which satisfies the vacuum Einstein equations with a cosmological constant  

\beq
\Lambda = - {d(d-1) \over 2 r_0^2}~.
\eeq  

\vspace{0.3cm}

\noindent
3) The {\it Robertson-Walker} space\footnote{After the american and british mathematicians, who in 1930s showed 
the generality of this space.} which corresponds to an homogeneous 
and isotropic spacetime. 
If $(M_4, g)$ is a $4$-dimensional manifold of constant curvature, 
corresponding to Euclidean ${\bf R}^{3}$ ($k=0$),
spherical ${\bf S}^{3}$ ($k=1$), or hyperbolic ${\bf H}^{3}$
($k=-1$) spaces or quotients of these by 
discrete groups of isometries, then $M_4 = {\bf R} \times M_3$ and 

\begin{equation}
\label{RWmetric} ds^2 = -dt^2 + a^2 (t) \left [ d \chi^2 +
f^2(\chi) (d\theta^2 + \sin^2 \theta  d\phi^2) \right ] \;,
\end{equation}
where $f(\chi)=(\chi\,$, $\sin\chi$, or $\sinh\chi)$, depending on
the value of the constant spatial curvature ($k=0,1,-1$). This metric 
is a solution of the Einstein equations for matter that can be 
described as a perfect fluid with velocity ${\bf u} = {\partial \over \partial t}$ and energy density 

\begin{equation}
\label{density} \rho = {\rho_0 \over a^{\gamma}} ~,
\end{equation}
where $\rho_0$ is a constant, $\gamma= (3)4$ corresponds to (non-)relativistic 
matter and the scale factor, $a(t)$, satisfies the Friedmann equation, 
a constraint equation,

\begin{equation}
{\dot a^2\over 2} - {4 \pi G \rho_0 \over 3 a^{\gamma-2}} = -{k \over 2} ~,
\label{Friedmann}
\end{equation}
which can be easily recognized as the first integral of motion of a unit mass 
particle in the potential $V(a) = - {4 \pi G \rho_0/ 3 a^{\gamma-2}}$.

Notice that, since $\rho_0$ is positive, then so is the energy density and hence the scale factor blows up in a finite time, 
which corresponds to a curvature singularity, the {\it Big Bang} or the {\it Big Crunch}. This means that time-like geodesics 
of the integral curves of ${\partial \over \partial t}$ are incomplete. Actually, this is a fairly general feature of 
spacetime, a result known as Hawking-Penrose singularity theorem, according to which physically meaningful Lorentzian 
manifolds are singular, i.e. are geodesically incomplete. A pedagogical and comprehensive 
introduction to the simplest singularity theorem of Hawking and Penrose can be found in 
Ref. \cite{Natario}.

\noindent 
It is relevant to point out that an important condition in the Hawking-Penrose singularity theorem is 
the one which concerns the physical nature of a manifold. A Lorentzian manifold $(M, g)$ is said to be 
physically reasonable when it satisfies the {\it strong energy condition}:

\begin{equation}
\label{Strongenergy} R_{\mu \nu} V^{\mu} V^{\nu} \ge 0~,
\end{equation}  
for any timelike vector field, $V^{\mu}$. From Einstein's equations this statement is equivalent, 
for $d \ge 2$, to the condition on the energy-momentum tensor and its trace, $T$, 

\begin{equation}
\label{EMtensor-strong} T_{\mu \nu} V^{\mu} V^{\nu} \ge {T \over d-1}  V_{\mu} V^{\mu}~, 
\end{equation}  
which is satisfied by the vacuum, the cosmological constant, if $\Lambda \ge 0$, and by a perfect fluid if $\rho + 3 p \ge 0$. 
Note that this condition is not respected during the inflationary period and by the present state of the Universe.

Another important feature of spacetime concerns its topology. 
Locally, a topology is induced by the distance function $d(P,Q)$ between points $P$ and $Q$ in ${\bf R}^d$ 
through the definition of open sets, that is, sets for which $d(P,Q) < r_0$, where $r_0$ is an arbitrary quantity. 
The properties of open sets follow from the Hausdorff's condition or separation axiom, according to which points $P$ and 
$Q$ in ${\bf R}^{d}$ have non-intersecting neighborhoods $U$ and $V$ such that $U \owns P$ and $V \owns Q$. It follows that 
the intersection of open sets is an open set and that the union of any number of open sets is also an open set. 
Topology also concerns the global structure of a space and can be classified by the differential forms it admits. 
The topology of low dimensional spaces ($d \le 3$) is fully characterized by its genus.  

It is rather remarkable that, on the largest scale, spacetime can be modeled by a $4$-dimensional manifold 
$M_4$ which is decomposed into
$M_4 = {\bf R} \times M_3$, and is endowed with a locally
homogeneous and isotropic Robertson--Walker metric, Eq.~(\ref{RWmetric}).
As we have seen, the spatial section $M_3$ is often taken to be one of the following
simply-connected spaces: Euclidean ${\bf R}^{3}$,
spherical ${\bf S}^{3}$, or hyperbolic ${\bf H}^{3}$ spaces. However,
$M_3$ may be actually a multiply connected quotient
manifold $M_3 = \widetilde{M}/\Gamma$, where $\Gamma$ is a fixed
point freely acting group of isometries of the covering space
$\widetilde{M}=({\bf R}^{3},{\bf S}^{3}, {\bf H}^{3})$. 

It is known that for the Euclidean geometry, besides
${\bf R}^{3}$  there are 10 classes of topologically distinct
compact $3$-dimensional spaces consistent with this geometry, while for the
spherical and hyperbolic geometries there are actually an
infinite number of topologically inequivalent compact manifolds
with non-trivial topology \cite{Luminet}.

It is no less remarkable that the spacetime topology can, at least in principle, be tested via the study of multiple images 
in the Cosmic Microwave Background Radiation (CMBR).
A quite direct strategy to test the putative non-trivial topology of the spatial sections of the Universe 
is the so-called ``circles-in-the-sky" method. It relies on the search of
multiple images of correlated circles in the CMBR maps~\cite{CSS1998}. 
Thus, in a non-trivial topology, the sphere of last
scattering intersects a particular set of images along pairs of
circles of equal radii, centered at different points on the last
scattering sphere with the same distribution 
of temperature fluctuations. 

It has been argued that an important evidence for a non-trivial topology 
arises from the fact that the Poincar\'e dodecahedral and
the binary octahedral spaces can account for the observed low value
of the CMBR quadrupole and octopole moments
measured by the WMAP team \cite{Poincare,Aurich,WMAP-Spergel}. 
However, a more recent search 
for the circles-in-the-sky, down to apertures of about $5^o$ using WMAP three years data has not been successful in 
confirming this possibility \cite{Key}.  

Of course, a topologically non-trivial space can be only detected if the Universe is not 
exceedingly larger than the size of the last scattering surface, which is clearly a quite restrictive 
condition and consistent with a rather modest period of inflation in the early Universe.
Even though, it is worth mentioning that through the circles-in-the-sky method one can obtain, besides the constraints 
arising from the usual astrophysical observational methods (supernova, baryon acoustic oscillations, 
CMBR bounds, etc), additional limits to the cosmological models. This can be shown to be 
particularly relevant for the $\Lambda$CDM model \cite{RAMM}, for the unified model of 
dark energy and dark matter, the Generalized Chaplygin Gas model \cite{Bento2006a}, 
characterized by the equation of state $p= - A/ \rho^{\alpha}$, where $p$ is the pressure, $\rho$, the energy 
density and $A$ and $\alpha$ are positive constants \cite{Bento2002a}, and for modified gravity models 
inspired in braneworld constructions \cite{Bento2006b}.

Another relevant issue about the property of spaces concerns their boundaries.
In $d=4$ the theory of {\it cobordism} guarantees that for all
compact 3-surfaces there always exists a compact 4-dimensional
manifold such that $S^3$ is the only boundary, or equivalently,
all 3-dimensional {\it compact} hypersurfaces are cobordant to zero
\cite{Stong}.  This question is particularly relevant 
when considering the sum of 
histories in Quantum Cosmology. In these approach, 
the quantum state of a $d=4$ Universe is described by a
wave function $\Psi[h_{ij}, \Phi]$, which is a functional of the
spatial 3-metric, $h_{ij}$, and matter fields generically denoted by
$\Phi$ on a compact 3-dimensional hypersurface $\Sigma$. The
hypersurface $\Sigma$ is then regarded as the boundary of a compact
4-manifold $M_4$ on which the 4-metric $g_{\mu\nu}$ and the
matter fields $\Phi$ are regular. The metric $g_{\mu\nu}$ and the
fields $\Phi$ coincide with $h_{ij}$ and $\Phi_0$ on $\Sigma$ and the
wave function is then defined through the path integral over
4-metrics, $^{4}g$, and matter fields:
\begin{equation}
\Psi[h_{ij}, \Phi_0] = \int_{{\cal C}} D[^{4}g] D[\Phi]
\exp\left(-S_{E}[^{4}g, \Phi]\right) ~,
\label{eq:1.1}
\end{equation}
where $S_E$ is the Euclidean action and $\cal C$ is the class of
4-metrics and regular fields $\Phi$ defined on Euclidean
compact manifolds and with {\em no other} boundary than
$\Sigma$. This wave function is the solution of the Wheeler-DeWitt equation, and it has been 
argued that this formalism allows for a theory of the initial conditions for the Universe \cite{HHawking}. Indeed, 
in this proposal, 
the wave functions are associated with a probability distribution and the most likely observational features of the Universe 
correspond to the peak of the solution of the Wheeler-DeWitt equation.     
Explicit solutions of this equation in the so-called {\it minisuperspace} approximation are known 
for some cases of interest, such as for a universe dominated by a massless 
conformally coupled scalar field \cite{HHawking} and by radiation \cite{BMourao}.

An extension of this proposal for universes
with $d > 4$ dimensions has some complications. In these
$d$-dimensional models, the wave function would be a functional of the
$(d-1)$ spatial metric, $h_{IJ}$, and matter fields, $\Phi$, on a
($d-1$)-hypersurface, $\Sigma_{d-1}$, and is defined as the result of
performing a path integral over all compact $d$-metrics and regular
matter fields on $M_{d}$, that match $h_{IJ}$ and the matter fields on
$\Sigma_{d-1}$.
One starts assuming that the $(d-1)$-surface $\Sigma_{d-1}$
does not possess any disconnected parts. Is there always
a $d$-dimensional manifold $M_d$ such that $\Sigma_{d-1}$ is
the only boundary ?  In higher dimensional manifolds, this is actually 
not guaranteed. There exist compact $(d-1)$-hypersurfaces
$\Sigma_{d-1}$ for which there is no compact $d$-dimensional manifold
such that $\Sigma_{d-1}$ is the only boundary.  This seems to indicate
that in $d > 4$ dimensions there are configurations which cannot be
attained by the sum over histories in the path integral.  The wave
function for such configurations would therefore be zero. However, 
this situation can be circumvented so as to obtain
non-vanishing wave-functions for such configurations, namely by dropping
the assumption that the $(d-1)$-surface $\Sigma_{d-1}$ does not
possess any disconnected parts (see {\it e.g.} \cite{Halliwell,BFM} and references 
therein).

Indeed, if one assumes that the hypersurfaces
$\Sigma_{d-1}$ consist of any number $n>1$ of disconnected parts
$\Sigma_{d-1}^{(n)}$, then one finds that the path integral for this
disconnected configuration involves terms of two types. The first type
consists of disconnected $d-$manifolds, each disconnected part of
which closes off the $\Sigma^{(n)}_{d-1}$ surfaces separately. These
will exist only if each of the $\Sigma^{(n)}_{d-1}$ are cobordant to
zero, but this may not always be the case. There will indeed be a
second type of term which consists of connected $d$-manifolds joining 
some of the $\Sigma_{d-1}^{(n)}$ together.  This
second type of manifold will always exist in any number of
dimensions, providing the $\Sigma_{d-1}^{(n)}$ are 
topologically similar, {\it i.e.} have the same {\it characteristic numbers}. 
The wave function of any $\Sigma_{d-1}^{(1)}$ surface
which is not cobordant to zero would be non-vanishing and can be 
obtained by assuming the existence of other surfaces of suitable
topology and then summing over all compact $d-$manifolds which join
these surfaces together.  Thus, given a compact $(d-1)$ hypersurface
$\Sigma_{d-1}$ which is not cobordant to zero, a non-zero amplitude
can be found by assuming it possesses disconnected parts.

However, the above considerations for disconnected pieces and generic
$\Sigma_{d-1}$ surfaces would spoil the Hartle-Hawking prescription,
since the manifold would have more than one boundary.  In other words,
the general extension discussed above would imply in a description in
terms of propagation between such generic $\Sigma_{d-1}$ surfaces.
The wave function would then depend on every piece and not on a single
one.  Nevertheless, if one restricts oneself
to the case of a truncated
model with a global topology given by a product of a 3-dimensional
manifold to a $d$-dimensional one, then the spacelike sections always
form a boundary of a $d$-dimensional manifold with no other boundaries. 
Since hypersurfaces $S^3 \times S^d$ are always cobordant
to zero, it implies that for spacetimes with topology ${\bf R} \times
S^3 \times S^d$ the Hartle-Hawking proposal can be 
always implemented \cite{BFM}. 

Let us close this section, introducing a notion that has been recently 
quite useful in physics, namely the idea of an {\it orbifold}. 
From the mathematical point of view, an orbifold is a generalization of the concept  
of manifold which includes the presence of the points whose neighborhood is 
diffeomorphic to the coset ${\bf R}^d/ \Gamma$, where $\Gamma$ is a 
finite group of isometries. In physics, an orbifold usually describes an 
object that can be globally written as a coset $M/G$, where $G$ is the group of its isometries 
or symmetries. 
The best known case of an orbifold corresponds to a manifold with boundary since it carries a natural orbifold structure, 
the ${\bf Z}_2$-factor of its double. Thus, a factor space of a manifold along a smooth ${\bf S}^1$-action without fixed 
points carries an orbifold structure.

In what follows we shall describe the properties that physical theories require for the physical spacetime.

\section{Physical Spacetime}
\label{sec:3}

Within the framework of General Relativity, the dynamics of the physical spacetime is actually related with the history
and evolution of the Universe.
The mathematical description of spacetime does allow for a wide range of scenarios; however, recent 
developments in observational cosmology do indicate that our Universe is well described by a flat Robertson-Walker 
metric, meaning that the energy density of the Universe is fairly close to the critical one, 
$\rho_c \equiv 3H_0^2/8 \pi G \simeq 10^{-29} g/cm^3$, 
where $H_0 \simeq 73~km~s^{-1}Mpc^{-1}$ is the Hubble expansion parameter at present. 
Furthermore, CMBR, Supernova and large scale structure 
data are consistent with 
each other if and only if the Universe is dominated by a smoothly distributed energy that does not manifest 
itself in the electromagnetic spectrum - dark energy. Moreover, it is found that the large scale structure of the Universe, as
well as the dynamics of galaxies, requires matter that like dark energy, does not manifest electromagnetically - dark matter. 
More exactly, in the cosmic budget of energy, dark energy corresponds to about $73\%$ of the critical density, 
while dark matter to about $23\%$ and baryonic matter, the matter that we are made of, to only about $4\%$ \cite{WMAP3}.  

Actually, the dominance of dark energy at the present does have deep implications for the evolution of spacetime. 
For instance, if dark 
energy remains the dominant component in the energy budget in the future, then 
geometry is no longer the determinant factor in the destiny of the Universe. As is well known, 
in a Universe where dark energy is sub-dominant, 
flat and hyperbolic geometries give origin to infinity universes in the future; in opposition, a spherical universe does 
eventually recollapse and undergoes 
a Big Crunch in a finite time. If however, dark energy is the dominant component, the fate of the Universe is determined 
by the way it evolves. If its energy density is decreasing, the Universe will eventually be dominated by matter and its destiny 
is again ruled by its geometry as described above. If, on the other hand, the energy density remains constant, 
then the Universe expansion will continue to grow and the universe will be quite diluted of matter. That is to say that, 
in the remote future the Universe will correspond to a dS  space with a future horizon. 
This means that the world will have features similar to an isolated thermal cavity with finite temperature and 
entropy. A more drastic fate is expected if 
the energy density of dark energy continues to grow. This growth will eventually cause a {\it Big Rip}, that is, 
the growing velocity of the spacetime expansion will eventually disrupt its very fabric and all known structures will 
be ripped off.  

Actually, an ever accelerating universe might not be compatible with some fundamental physical theories. For instance, 
an eternally accelerating universe poses a challenge for string theory, at least in
its present formulation, as it requires that its asymptotic states are asymptotically free, which is 
inconsistent with a spacetime that exhibits future horizons \cite{Hellerman,Fischler,Witten2}. 
Furthermore, it is pointed out that theories with a stable supersymmetric
vacuum cannot relax into a zero-energy ground state if the
accelerating  dynamics is guided by a single scalar field
\cite{Hellerman,Fischler}. This suggests that the accelerated expansion 
might be driven by at least two scalar fields. 
It is interesting that some two-field models allow for solutions with an exit from a period of accelerated
expansion, implying that decelerated expansion is resumed (see {\it e.g.} Ref. \cite{Bento2002b}). 
Hence, a logical way out of this  
problem is to argue that the dS space is unstable. This might also occur, for instance, due to quantum 
tunneling, if the cosmological constant is not too small.  
  
Another significant feature about our Universe is that only if it has undergone a period of quite rapid and accelerated expansion 
in its early history, one can understand why its spatial section is so close to flat and why it 
is so homogeneous and isotropic on large scale \cite{Guth,Linde,ASteinhardt}. 
This {\it inflationary} phase of accelerated expansion, a tiny fraction of a second after the Big Bang, 
about $10^{-35}$ seconds, corresponds to a period where the geometry of the Universe is described by a 
dS space. It is quite remarkable that a rather brief period of inflation, a quite generic behavior of most  
of the anisotropic Bianchi-type spaces \cite{Wald}, Kantowski-Sachs spaces \cite{TWidrow} and inhomogeneous spaces 
\cite{PWilliams} dominated 
by a cosmological constant, drives a microscopic universe into a large one, whose features closely resemble ours. 
Moreover, in 
this process, small quantum fluctuations of the field responsible for inflation, the {\it inflaton}, are amplified to 
macroscopic sizes and are ultimately responsible for the formation of large scale structure (see {\it eg.} Ref. \cite{Olive} for 
an extensive discussion). It is a great achievement of 
modern cosmology that the broad lines of this mechanism are corroborated by the observed features of the CMBR, 
such as its main peak, whose position is consistent with the size of the scalar 
density fluctuations that first reentered the horizon, as well as the nearly scale invariant and Gaussian nature of these 
fluctuations.   

However, in what concerns spacetime, the stock of surprises arising from physics is far from over. Indeed, recent developments 
on the understanding of string theory have led to speculations that may be regarded as somewhat disturbing for those who 
believe that the laws of nature can be described by an {\it action}, which encompasses the relevant 
underlying fundamental symmetries, and from which an unique vacuum arises and the spectrum of elementary objects, particles, can be 
found. These view has been recently challenged by a quite radical set of ideas. The genesis of these can be traced from 
the understanding that the initial outlook concerning the original 
five distinct string theories was not quite correct. It is now understood that there is instead 
a continuum of theories, that includes M-theory, interpolating the original five string theories. One rather speaks of 
different solutions of a master theory than of different theories. The space of these solutions is often referred to as the 
{\it moduli space of supersymmetric vacua} or {\it supermoduli-space}. These moduli are fields, and their variation allows 
moving in the supermoduli-space. The moduli vary as one moves in the spacetime, as moduli have their own equations of motion.

However, the continuum of solutions in the supermoduli-space are supersymmetric and have all a vanishing cosmological constant. 
Hence, in order to describe our world, there must exist some non-supersymmetric ``islands'' in the supermoduli-space. 
It is believed that the number of these discrete vacua is huge, googles, $G = 10^{100}$, or googleplexes $10^G$, 
instead of unique \cite{Susskind1}. If the cause of the accelerated expansion of the Universe is due to a small cosmological constant, 
then the state of our Universe corresponds to moduli values some of the non-supersymmetric islands 
in the supermoduli-space. The fact that the magnitude of the 
cosmological constant is about $10^{120}$ smaller than its natural value $M_P^4$, where $M_P= 1.2 \times 10^{19}$ GeV 
is the Planck mass, makes it highly unlikely to find such a vacuum, unless there exists a huge number of solutions with 
every possible value for the cosmological constant. 
The space of all such string theory vacua is often referred to as the {\it landscape} 
\cite{BoussoPolchinski00}. 

From the landscape proposal springs a radical scenario. In principle, vacua of the landscape do not need to correspond to 
actual worlds, however, very much on the contrary, it is argued that the string landscape 
suggests a {\it multiuniverse}. According to this proposal, the multiple vacua of string theory is associated 
to a vast number of ``pocket 
universes'' in a single large Mega-universe. These pocket universes, like the expanding universe we observe around us, are all 
beyond any observational capability, as they lie beyond the cosmological horizon. In the words of Susskind, 
a vociferous proponent 
of the multiuniverse idea \cite{Susskind2}, ``According to classical physics, those other worlds are forever completely sealed 
off from our world''. Clearly, the implications of these ideas are somewhat disturbing. 
First, the vacuum that corresponds to our world must arise essentially form a selection procedure, 
to be dealt with via anthropic or quantum cosmological considerations. Thus, it seems that somehow our existence plays 
an important role in the selection process. Second, the vast number of vacua in the landscape ensures 
the reality of our existence; one refers to this scenario as the {\it anthropic landscape}, when based on anthropic arguments. 
For sure, this interpretation is not free from criticism. It has been pointed out, for instance, 
that the impossibility of observing a multiuniverse implies that its scientific status is 
questionable. It is in the realm of metaphysics, rather than of physics \cite{GEllis06}. It has also been argued that selection 
criteria like the anthropic landscape must be necessarily supplemented by arguments based on dynamics and symmetry, 
as only these lead to a real ``enlightenment'', the former are actually 
a ``temptation'' \cite{Wilczek05}.   
Indeed, Weinberg argues that the anthropic reasoning makes sense for a given constant whenever the range over which 
it varies is large compared with the anthropic allowed range. That is to say, it is relevant to know what 
constants actually ``scan''. The most likely include the cosmological constant, and the particle masses set by 
the electroweak symmetry breaking mechanism. The possibility that the later is anthropically fixed is regarded as 
an interesting possibility, given that it renders an alternative solution for the hierarchy problem, 
such as Technicolor or low-energy supersymmetry, that are not fully free of problems \cite{Weinberg05}. 
In any case, we feel that we cannot close this discussion without some words of caution. For instance, 
Polchinski has recently pointed   
out as the landscape picture requires a higher level of theoretical skepticism given that it suggests 
that science is less predictive. Furthermore, he remarks that the current scenario is tentative 
at best, as a nonperturbative formulation of string theory is still missing \cite{Polchinski06}.  

Let us close this discussion with a couple of remarks. The first concerns the possibility 
that the topology of the landscape is non-trivial. This hypothesis would imply 
that multiuniverses are not causally exclusive, meaning 
that within our Universe one might observe pocket sub-universes 
where the laws of physics are quite different from the ones we know. Since it is natural to assume the in 
these sub-universes the fundamental constants assume 
widely different values, one might expect to observe oddities such as quantum phenomena on macroscopic scales 
and relativistic effects at quite mundane velocities. Of course, this possibility would imply in a further loss of 
the ability to predict the properties of the cosmos. 

Another relevant investigation on the selection of landscape vacua concerns the understanding of 
up to which extent the problem can be 
addressed in the context of the quantum cosmology formalism. As already discussed, this formalism allows for a theory of 
initial conditions, which seems to be particularly suitable to deal with the problem of vacua selection. It is quite interesting 
that this problem can be addressed as an N-body problem for the multiple scattering among the N-vacua sites of the landscape 
\cite{HMersini}. The use of the Random Matrix Theory methods shows that the phenomenon of localization on a lattice site with 
a well defined vacuum energy, the so-called Anderson localization, occurs. It is found that the most probable universe with 
broken supersymmetry corresponds to a dS universe with a small cosmological constant.  
Furthermore, it is argued that the relevant question on why the Universe started in a low entropy state can only be 
understood via the interplay between matter and gravitational degrees of freedom and the inclusion of dynamical 
back-reaction effects from massive long wavelength modes \cite{HMersini}. It is interesting to speculate whether these 
features remain valid beyond scalar field models, the case considered in Refs. \cite{HMersini}. Massive vector fields with 
global $U(1)$ and $SO(3)$ symmetries seem to be particularly suitable to generalize these results, given that the 
reduced matrix density and Wigner functional of the corresponding {\it midisuperspace} model \cite{BMoniz} exhibit properties 
that closely resemble the localization process induced by the back-reaction of the massive 
long wavelength modes discussed in \cite{HMersini}.   

Let us describe some recent developments 
involving the AdS space, introduced in the previous section. 

In the so-called {\it braneworlds}, one can admit  
two $3$-branes at fixed positions along the 5th dimension, such that 
the {\it bulk}, the $5$-dimensional spacetime is AdS, with a negative cosmological 
constant, $\Lambda = - 3 M_5^3 k^2$, where $M_5$ is the $5$-dimensional Planck mass and $k$ a constant with 
dimension of mass. In this setup, 
compactification takes place on a ${\bf S^1/Z^2}$ orbifold symmetry. 
Einstein equations admit a solution that preserves Poincar\'e 
invariance on the brane, and whose spatial background has a non-factorisable geometry with an 
exponential warp form

\beq
ds^2 = e^{-2k \vert z \vert} g_{\mu \nu} dx^{\mu}dx^{\nu} + dz^2 ~,
\eeq
$g_{\mu \nu}$ being the $4$-dimensional metric.

The action of the model is given by \cite{RandallSundrum99}

\beq 
S_5 = 2 \int d^{4}x \int_{0}^{z_{c}} dz \sqrt{-g_5} \left[{M_5^3 \over 2} 
R_{5} - 2 \Lambda\right] - 
\sigma_{+} \int d^{4}x \sqrt{-g_{+}} -  
\sigma_{-} \int d^{4}x \sqrt{-g_{-}} ~,
\eeq
corresponding to the bulk space with metric, $g_{5MN}$, a  
with a positive tension, $\sigma_{+}$, brane with metric, $g_{+\mu \nu}$, sitting at $z = 0$, 
and a negative tension, $\sigma_{-}$, brane with metric, $g_{-\mu \nu}$,
sitting at $z = z_{c}$. The Standard Model (SM) degrees of freedom lie presumably on the brane at $z = z_{c}$.

In order to ensure a vanishing cosmological constant in $4$-dimensions one chooses: 

\beq
\sigma_{+} = - \sigma_{-} = 3 M_5^3 k ~.
\eeq

This is quite interesting, as a cancellation involving different dimensions is actually 
not possible within the context of the Kaluza-Klein compactification mechanism \cite{Rubakov}.  
Another pleasing feature of this proposal is that the hierarchy between the Planck and the SM scales can be 
dealt with in a geometrical way, actually via the warp $e^{-2k z_{c}}$ factor. 
Finally, integration over ${\bf S^{1}}$ allows obtaining Planck's constant in $4$-dimensions:

\beq
M_{P}^2 = {M_5^3 (1 - e^{-2k z_{c}}) \over 4 k} ~~,
\eeq
which clearly exhibits a low dependence on $kz_{c}$.

The literature on braneworlds is quite vast and it is not our aim to review here the most important proposals, however it is 
interesting to mention that the type of cancellation mechanism described above can be also considered to understand why Lorentz 
invariance is such a good symmetry of nature in $4$ dimensions \cite{BertolamiCarvalho06}. 
Furthermore, it is worth mentioning that evolving 
$3$-branes can be regarded as solutions of an effective theory that arises from the fundamental M-theory. Indeed, the  
$d=11$ M-theory, compactified on an ${\bf S^1/Z^2}$
orbifold symmetry with $E_8$ gauge multiplets on each of the $10$-dimensional 
orbifold fixed planes, can be identified with the strongly coupled $E_{8} \otimes E_{8}$ heterotic theory \cite{HorWitten96}. 
An effective theory can be constructed via the the reduction of
the $d=11$ theory on a Calabi-Yau three-fold space, ${\bf K}$, that is,
$M_{11} = {\bf R}^{4} \times {\bf K} \times {\bf S^1/Z^2}$. It is shown  that this effective theory admits  
evolving {\it cosmological domain-wall solutions} corresponding to a pair of $3$-branes  \cite{LukasOvrutWaldram99}.

Furthermore, it is relevant to realize that an AdS space is the natural background for 
supergravity and M-theory, given that in the weak coupled limit 
the latter theory corresponds to a $N=1$ supergravity theory in $11$-dimensions. 
Thus, the AdS space is intimately related with 
string theory. Actually, this background space is quite crucial in the so-called Maldacena or AdS/CFT conjecture 
(see Ref. \cite{Aharony} for an extensive discussion), according to
which a supergravity theory in $d$-dimensions on a AdS space is equivalent to a conformal field theory (CFT) defined 
on the $(d-1)$-dimensional boundary of that theory.

Before drawing an end to our brief discussion on some of the properties of 
spacetime, let us discuss one last striking development concerning the nature of spacetime. 
It has been suggested that at the most fundamental level, the underlying geometry of spacetime is  
noncommutative. This feature arises from the discovery in string theory that the low-energy effective theory of a D-brane 
in the background of a NS-NS B field lives in a noncommutative space \cite{Connes,Seiberg,Schomerus} where the configuration 
variables satisfy the commutation relation:

\begin{equation}
\label{commutator}
\left[ x^{\mu },x^{\nu }\right] =i\theta ^{\mu \nu }.
\end{equation}
where $\theta ^{\mu \nu }$ is a constant antisymmetric matrix. Naturally, this 
set of numbers do not transform covariantly, which implies
in the breaking of Lorentz invariance down to the stability subgroup of the 
noncommutative parameter \cite{Carroll}. Approaches where $\theta ^{\mu \nu }$ is regarded as a 
Lorentz tensor were considered, for instance, in the context of a noncommmutative scalar field coupled
to gravity in homogeneous and isotropic spaces \cite{BGuisado1}.   
 
Naturally, if spacetime has a noncommutative structure one should expect important implications in field theory 
(see Refs. \cite{Szabo,Doug_Nekra} for extensive discussions) and even in the non-relativistic limit, that is, 
at Quantum Mechanics level. In the first case, one finds a host 
of new effects including the violation of translational invariance (see \cite{BGuisado2} and references therein). 
In the non-relativistic limit, versions of noncommutative Quantum Mechanics (NCQM) have been recently the subject of many studies.  
Although in string theory only the space coordinates exhibit a noncommutative structure, some authors have suggested 
NCQM models in which noncommutative geometry is defined in the whole phase space \cite{Zhang_1,Bertolami_1,Djemai}.  
Implications for the gravitational quantum well, recently realized for ultra-cold neutrons from the research reactor of the
Laue-Langevin Institute in Grenoble \cite{Nesvizhesky}, have been examined for the NCQM models with a phase 
space noncommutative geometry \cite{Bertolami_1,Bertolami_2}.

\section{Concluding remarks}
\label{sec:4}

Physics has unquestionably made untenable the philosophical thinking according to which 
space and time are {\it a priori} concepts, independent 
of the physical world. Physics has also immensely stretched the notions of space and time, expanding reality to limits 
that were thought to be beyond imagination. 
Indeed, the physical world was, according to Aristotle, compact and locked within the sub-lunar realm.     
Galileo's observations and the universality of Newton's mechanics have fundamentally changed that. 
XIX century physics was rather modest about the timescale of the world, based on thermodynamical considerations about 
dissipation of heat 
and the conversion of gravitational energy into heat. Indeed, estimates by Lord Kelvin and Helmholtz suggested a few hundred millions 
of years for the ages of the Sun and Earth. Geologists were actually the first to understand that 
this could not be possible. Earth had to be at least a billion years old to be consistent with the transformations that 
are in operation at present. Paleontologists followed suit, 
given the tight correlation between fossils and the geological strata they are found. On its hand, 
astronomy has open up space and time, providing 
us with impressive estimates of the size and the distance of astronomical objects, having ultimately 
shown us that space itself 
is expanding - in fact in an accelerated fashion according to the most recent observations.   
The ticket to fully exert the
freedom to expand space and time was conquered when Einstein understood that General Relativity 
was a theory of the spacetime at large. Since then, 
scrutinizing the ways spacetime might exist is, in a 
way, the very essence of physics. Physics has thus given substance to the pioneering work of scores of brilliant mathematicians 
who speculated on the geometry and topology of spaces. 

According to the swiss painter, Paul Klee, ``L'art ne reproduit pas le visible, il rendre le visible'', 
and, in a broader sense, the same can be said about physics. Indeed, from its original goal of describing nature, physics 
has created a picture of the world that is much richer than the one that meets the eye, and it turns out
that, in this process, spacetime has acquired a quite rich structure. However, the adventure is by no means over. 
On a quite fundamental level,  
we do not understand how to reconcile our picture of the macroscopic spacetime with the rules of Quantum Mechanics, a
theory that successfully describes all, but gravitational phenomena. This is an 
unbearable gap in our knowledge. Moreover, this difficulty has quite 
severe and concrete implications, the most evident being, as we have seen, that we cannot explain the 
smallness of the cosmological constant without paying a quite heavy toll. In fact, the cosmological 
constant problem is such a formidable challenge that it is 
tempting to go around it and compare it with Wittggenstein's suggestion, according to which 
all problems of philosophy are actually problems of language. Indeed, our expectation that the 
cosmological constant is immensely 
greater than the observed value on cosmological scales is based on the ``language'' of quantum 
field theory. We do not expect and 
we have not seen a breakdown of the quantum field theory formalism down to scales of about $10^{-18}$ m, but this still  
a long away from the typical quantum 
gravity length scale,  $L_P \simeq 10^{-35}$ m. In fact, it is relevant to bear in mind that the cosmological constant 
problem is intimately related with supersymmetry, duality symmetries and the spacetime dimensions \cite{Witten3}.  
When put together, these ingredients may imply in an important ``language'' shift, as is the case of the landscape 
scenario. Most likely, the ultimate landslide is still to come. In any case, unraveling the 
ultimate structure of spacetime down to the
smallest scale, and then back up to the largest one will remain, as it is nowadays, an exciting quest for many generations to 
come.  

%
%
%



\printindex
\end{document}